\documentstyle[12pt,damtp]{damtpprint}
\pubdate{October 1991}
\pubnumber{DAMTP-91/35}
\title{Projections of Virasoro Singular Vectors}
\author{Adrian Kent\damtp}
\abstract{
We use recently derived explicit formulae for the Virasoro algebra's
singular vectors to give constructive proofs of three results due to
Feigin and Fuchs.
The main result, which is needed for a rigorous
treatment of fusion, describes the action of the
singular vectors on conformal fields.
}

\begin{document}
\maketitle

\input mssymb
\newenvironment{proof}{{\bf Proof}}{}
\newtheorem{lemma}{Lemma}
\newtheorem{theorem}[lemma]{Theorem}
\newtheorem{corollary}[lemma]{Corollary}
\def\floor#1{\lfloor#1\rfloor}
\def\ceil#1{\lceil#1\rceil}
\def\blank#1{}
\def\swap#1#2{#1}
\def \ul {\underline}
\def \eps {\epsilon}
\def\ut#1{\hbox{\boldmath #1}}
\def\cO{{\cal O}}
\def\cFlm{{\cal F}_{\lambda,\mu}}
\def\tr{{\rm tr}}
\def\romif{{\rm~if~}}
\def\sgn{{\rm sgn}}
\def\sig{{\rm sig}}
\def\diag{{\rm diag}}
\def\for{{\rm ~for~}}
\def\and{{\rm ~and~}}
\def\romor{{\rm ~or~ }}

\section{Introduction}

The simplest conformal field theories are built from a finite number of
degenerate representations of the Virasoro algebra, and the
recent advances in our understanding of two dimensional physics have
relied on developments in the theory of these representations.
The degenerate representations are those for which the corresponding Verma
module contains a singular vector; the fact that the singular
vector can consistently be set to zero implies that the correlation functions
of the theory satisfy certain differential equations, and hence that
the theory is completely solvable.

Explicit expressions for the singular vectors are needed for a variety
of conformal field theoretic
applications.\cite{bpz,mattis,langlands,bdfiz2,ak1}
Two partial results in this direction were obtained by
Feigin and Fuchs,\cite{ff3} who gave the leading terms in an expansion
of the singular vectors and --- more significantly --- described the
action of the singular vectors on general conformal fields.
Benoit and Saint-Aubin\cite{bsa} found
expressions for a sub-series of the singular vectors.
These miraculous-looking expressions were made intelligible by
Bauer et al.\cite{bdfiz2,bdfiz1}
These authors also
set out\cite{bdfiz2}
a simple set of recursion relations from which expressions for
the general singular vectors can be obtained.
We have recently found\cite{ak2} simple expressions for all the singular
vectors, written as products of operators that involve non-integer powers
of $L_{-1}$.

Feigin and Fuchs' two theorems have remained slightly mysterious, in that
the proofs\cite{ff3,lutsjuk} are not constructive: they
begin with conjectures for the
relevant expressions and then verify that these conjectures must be correct.
A partial explanation was given by Bauer et al., who found an elegant
direct derivation of Feigin and Fuchs' results for the
Benoit---Saint-Aubin subseries of singular vectors.
The main point of this note is to
extend the arguments given by Bauer et al. and give direct
proofs of the Feigin-Fuchs theorems; we also prove a result of Feigin and Fuchs
concerning an invariance of the singular vectors.

\section{Singular vectors}

We first recall some elementary results.
The Virasoro algebra has commutation relations
\eqn{vcrs}
\begin{array}{rcl}
{}~[L_m , L_n ] \, \, & = & (m - n ) L_{m+n} + \frac{m^3 - m}{12}
\delta_{m,-n} C \, , \\ ~[L_m , C ] \, \, & = & 0 \, .
\end{array}
\en
Its Verma modules, written $V(h,c)$, are irreducible representations which
contain a vector $\ket{h}$ such that
\eqn{hwt}
\begin{array}{rcll}
L_m \ket{ h } &=& 0 & \romif m>0 \, , \\ L_0 \ket{ h } &=& h \ket{h} \, , & \\
C \ket{h} &=& c \ket{h} \, . &
\end{array}
\en
We have the decomposition
\eq
V(h,c) = \bigoplus_{n = 0,1,2 \ldots}  V_{n} (h,c)  \, ,
\en
where the level $n$ space
$V_n (h,c ) $ is the eigenspace of $L_0$ with eigenvalue $(h+n)$.
A singular vector in $V(h,c)$ is a vector $v$, lying in some
$V_n (h,c)$ for $n>0$, with the property that
\eqn{sing}
L_m v = 0  \romif m>0 \, .
\en
It is well known\cite{ff3,kac,ff1,ff2} that there is a
singular vector at level $N$ in
$V(h,c)$ if and only if, for some positive integers $p$ and $q$ and complex
number $t$, we have $N = pq$, and
\eqn{hc}
\begin{array}{rclcl}
c &=& c(t) &=& 13 - 6 t - 6 t^{-1} \, , \\
h &=& h_{p,q}(t)   &=& {\displaystyle
  \frac{p^2 -1}{4} t - \frac{pq -1}{2} + \frac{q^2 -1}{4} t^{-1}} \, .
\end{array}
\en
The singular vector at level $N$, when it exists, is
unique up to scalar multiplication: given $p$ and $q$, we can write
the singular vector $v_{p,q}(t)$ as a continuous function of $t$.
In fact, we have $v_{p,q}(t) = O_{p,q} (t) \ket{h_{p,q}(t)}$,
where
\eq
O_{p,q} (t) = \sum_{|I| = pq} a^{p,q}_I (t) L_{-I} \, .
\en
Here we write $L_{-I} = L_{-i_1} \ldots L_{-i_n}$
and $|I| = i_1 + \ldots + i_n$;
the sum is over sequences
$I = \{ i_1 , \ldots , i_n \}$ of positive integers ordered so that
$i_1 \geq \ldots \geq i_n $;
the $a^{p,q}_I (t)$ depend polynomially on $t$ and $t^{-1}$;
by convention, we take the coefficient of $(L_{-1})^{pq}$ to be $1$.

Now we recall some of the results obtained by Bauer et al.\cite{bdfiz2,bdfiz1}
Set $c = c(t)$ and
$h = h_{p,1} (t) $.
Introduce a sequence of vectors $f_p , f_{p-1} , \ldots , f_1 , f_0$ in
the Verma module $V(h,c)$, where each vector $f_r$ is at level $r$,
$f_0$ is the vector $\ket{h_{p,1}(t)}$ and the remaining vectors
are determined as follows.
Define the $p$-dimensional vectors
\eq
\begin{array}{rcl}
\ul{f} & = & ( f_{p-1}, f_{p-2} , \ldots , f_0 )^T  \, , \\
\ul{F} & = & ( f_p , 0 , \ldots , 0 )^T \, .
\end{array}
\en
The $p$-dimensional vector space carries an irreducible $sl(2)$
representation, of spin $(p-1)/2$.  Define the matrices $E^{i,j}$ by
$(E^{i,j})_{k,l} = \delta_{i,k} \delta_{j,l}$ and let
\eq
\begin{array}{rcl}
J_- & = & { \displaystyle \sum_{r=1}^{p-1} E^{r+1,r} \, , } \\
J_+ & = & { \displaystyle \sum_{r=1}^{p-1} r(p-r) \, E^{r,r+1} \, , } \\
J_0 & = & {\displaystyle \sum_{r=1}^{p} (\frac{p+1}{2}-  r ) \, E^{r,r} \, ,}
\end{array}
\en
so that
\eq
[ J_+ , J_- ] = 2 \, J_0 \, , ~~~~~ [ J_0 , J_{\pm} ] = \pm J_{\pm} \, .
\en
Now define the vectors $f_r$ for $1 \leq r \leq p$ by the equation
\eqn{bau1}
\ul{F} = ( - J_- + \sum_{k=0}^{p-1} L_{-k-1} (-t J_+ )^k ) \ul{f} \, .
\en
It can be shown\cite{bdfiz2} that the vectors obey
\eq
\begin{array}{rcl}
L_1 f_r & = & {\displaystyle
r (p - r) (-t(r - \frac{p+3}{2} ) -1) f_{r-1}} \, \\
L_2 f_r & = & {\displaystyle
r (p-r) (t^2 ( r - \frac{p}{2} - 3) - \frac{7}{4} t )
( -r^2 + (p+2) r - (p+1))  f_{r-2} } \,
\end{array}
\en
which in particular implies that
\eq
L_m f_p = 0 \, \for m > 0 \, .
\en
Hence $f_p$ is the singular vector $v_{p,1}(t)$.

Similarly, if we define
$f_0 = \ket{h_{1,q}(t)}$ and the $q$-dimensional vectors
\eq
\begin{array}{rcl}
\ul{f} & = & ( f_{q-1}, f_{q-2} , \ldots , f_0 )^T  \, , \\
\ul{F} & = & ( f_q , 0 , \ldots , 0 )^T \, ,
\end{array}
\en
and take the $q$-dimensional representation of $sl(2)$, then
the singular vector $v_{1,q}(t)$ can be expressed as $f_q$, where
\eqn{bau2}
\ul{F} = ( - J_- + \sum_{k=0}^{q-1} L_{-k-1} (-t^{-1} J_+ )^k ) \ul{f} \, .
\en

Eliminating the non-singular vectors in equations (\ref{bau1}) and
(\ref{bau2}), we find
\eqn{bsaops}
\begin{array}{rcl}
O_{p,1} (t) &=&
{\displaystyle \sum_{{I = \{ i_1 , \ldots , i_n \}}  \atop{ |I| = p}}}
c_p ( i_1 , \ldots , i_n ) (-t)^{p-n} L_{-I} \, , \\
O_{1,q} (t) &=&
{\displaystyle \sum_{{I = \{ i_1 , \ldots , i_n \}}  \atop{ |I| = q}}}
c_q ( i_1 , \ldots , i_n ) (-t)^{-q+n} L_{-I} \, ,
\end{array}
\en
where the sums are over all sequences of positive integers summing to $p$ or
$q$, without any ordering restriction, and the coefficients are defined by
\eq
c_r ( i_1 , \ldots , i_n ) =
{\displaystyle \prod_{{1 \leq k < r} \atop{k \neq \sum_{j=1}^s i_j
{\rm~for~any~}s}}} k (r - k ) \, .
\en
These are the expressions originally obtained by Benoit and
Saint-Aubin.\cite{bsa}

The general singular vectors can now be obtained as follows.\cite{ak2}
Firstly, we extend the Virasoro algebra by including operators of the
form $(L_{-1})^a$, for arbitrary complex $a$, with relations
\eqn{vtcrs}
\begin{array}{rcll}
{}~[L_m , ( L_{-1} )^a ] &=& {\displaystyle \sum_{n=1}^{m+1}}
({\displaystyle\prod_{r=1}^n}
\frac{(m+2-r)(a+1-r)}{r}) ( L_{-1} )^{a-n} L_{m-n}
& \romif m \geq 0 \, , \\
{}~[ (L_{-1} )^a , L_m ] &=& {\displaystyle \sum_{n=1}^{\infty}}
({\displaystyle\prod_{r=1}^n }
\frac{-(m+2-r)(a+1-r)}{r})  L_{m-n} (L_{-1})^{a-n}
                                         & \romif m < 0 \, ,
\end{array}
\en
and
\eq
(L_{-1})^a L_{-1} = L_{-1} (L_{-1})^a = (L_{-1})^{a+1} \, , ~~~
(L_{-1})^a (L_{-1})^b = (L_{-1})^{a+b}.
\en
Now we rewrite the expressions (\ref{bsaops}) by commuting all $L_{-1}$
operators to the right, with no other reordering.  This gives
\eq
\begin{array}{rcl}
O_{p,1} (t) &=& {\displaystyle
\sum_{r=0}^{\floor{\frac{p}{2}}}
\sum_{{ k_i \geq 2} \atop { \sum k_i \leq p}}}
P_{k_1, \ldots , k_r} (p,t) L_{- k_1} \ldots L_{- k_r}
(L_{-1})^{p - \sum k_i } \, , \\
O_{1,q} (t) &=& {\displaystyle
\sum_{r=0}^{\floor{\frac{q}{2}}}
\sum_{{k_i \geq 2} \atop { \sum k_i \leq q} }}
P_{ k_1, \ldots , k_r} (q, t^{-1}) L_{- k_1} \ldots L_{- k_r}
(L_{-1})^{q - \sum k_i }
\, ,
\end{array}
\en
where $P_{k_1, \ldots , k_r} (p , t)$ is defined for $p \geq \sum_i k_i$,
in which range it is a polynomial function in $p$ and $t$.
We analytically extend $P_{k_1, \ldots , k_r} (p , t)$ to arbitrary $p$ and
define operators
\eqn{anal}
\begin{array}{rcl}
\cO_{a,1} (t) &=& (L_{-1})^a + {\displaystyle \sum_{n=2}^{\infty}} \,
 {\displaystyle \sum_{{k_1 + \ldots + k_r = n} \atop {k_i \geq 2}}}
P_{ k_1, \ldots , k_r } (a, t) L_{- k_1} \ldots
L_{- k_r} (L_{-1})^{a - n } \, , \\
\cO_{1,b} (t) &=& (L_{-1})^b + {\displaystyle \sum_{n=2}^{\infty}} \,
 {\displaystyle \sum_{{k_1 + \ldots + k_r = n} \atop {k_i \geq 2}}}
P_{k_1, \ldots , k_r } (b , t^{-1}) L_{- k_1} \ldots
L_{- k_r} (L_{-1})^{b - n } \, ,
\end{array}
\en
for any complex numbers $a$ and $b$.

Then it can be shown\cite{ak2} that the following operator identities hold:
\eqn{soln1}
\begin{array}{rcl}
O_{p,q}(t) &=&
\cO_{p+(q-1)t^{-1},1} (t) \cO_{p+(q-3)t^{-1},1} (t) \ldots
\cO_{p-(q-1)t^{-1},1} (t) \\
&=& \cO_{1,q+(p-1)t} (t) \cO_{1,q+(p-3)t} (t) \ldots
\cO_{1,q-(p-1)t} (t) \, .
\end{array}
\en
These identities give two equivalent expressions for the singular vector:
\eqn{soln2}
\begin{array}{rcl}
v_{p,q}(t) & = & \cO_{p+(q-1)t^{-1},1} (t) \cO_{p+(q-3)t^{-1},1} (t) \ldots
\cO_{p-(q-1)t^{-1},1} (t) \ket{h_{p,q}(t)} \\
& = &
\cO_{1,q+(p-1)t} (t) \cO_{1,q+(p-3)t} (t) \ldots
\cO_{1,q-(p-1)t} (t) \ket{h_{p,q}(t)} \, .
\end{array}
\en
It can also be shown\cite{ak2} that
\eqn{soln3}
\begin{array}{rcl}
\cO_{p,1} (t) &=& O_{p,1} (t)\, , \\
\cO_{1,q} (t) &=& O_{1,q} (t)\, ,
\end{array}
\en
if $p$ and $q$ are positive integers,
and that
\eqn{soln4}
\begin{array}{rcl}
\cO_{-a,1} (t)\cO_{a,1} (t) &=& 1 \, , \\
\cO_{1,-b} (t)\cO_{1,b} (t) &=& 1 \, ,
\end{array}
\en
for any $a$ and $b$.

\section{Feigin-Fuchs' theorems}

The Virasoro algebra has a class of representations $\cFlm$, which are spanned
by vectors $w_r$ (for $r$ integer), and on which the algebra acts so that
$L_m$ and $C$ are represented by $\pi(L_m)$ and $\pi(C)$, with
\eqn{flm1}
\begin{array}{rcl}
\pi(L_m ) w_r & = & (\mu + r + \lambda ( m+1) ) w_{r-m} \, , \\
\pi(C) w_r & = & 0 \, .
\end{array}
\en
(Note that the parametrisation is redundant:
$\cFlm$ and ${\cal F}_{\lambda,\mu +1}$ describe the same representation.)
These representations describe the action of the Virasoro algebra on a
general conformal field.  For example, the Virasoro algebra's
adjoint representation is obtained by setting $\mu = - 1$, $\lambda = 1$,
$\pi (L_m )=  [ L_m , ~~ ]$ and $w_r = L_{-r}$.

The following results are due to Feigin and Fuchs.\cite{ff3}
Firstly, define the function $f_{p,q} ( \lambda , \mu , t)$ by the formula
\eq
\pi ( O_{p,q} (t) ) w_0 = f_{p,q} (\lambda , \mu , t) w_{pq}
\en
in $\cFlm$.
Let $\theta = \sqrt{-t^{-1}}$ and let
\eq
A_{p,q}(m,n) =
\left( (\frac{p-1}{2} + m )\theta^{-1} + (\frac{q-1}{2} + n ) \theta \right)
\left( (\frac{p+1}{2} - m) \theta^{-1} + (\frac{q+1}{2} - n) \theta  \right)
\, .
\en
Then
\eqn{ff1}
f_{p,q}^2 ( \lambda , \mu , t) =
{\displaystyle \prod_{{-\frac{p-1}{2} \leq m \leq \frac{p-1}{2}} \atop
                      {-\frac{q-1}{2} \leq n \leq \frac{q-1}{2}} }}
\left( (\mu + A_{p,q} (m,n))(\mu + A_{p,q} (-m,-n)) -
 4 \lambda (m \theta^{-1} + n \theta )^2 \right) \, ,
\en
where $m$ and $n$ range over
$-\frac{(p-1)}{2}, - \frac{(p-3)}{2} , \ldots , \frac{p-1}{2}$ and
$-\frac{(q-1)}{2}, - \frac{(q-3)}{2} , \ldots , \frac{q-1}{2}$
\linebreak respectively.

Secondly, let the operator $\sigma_{p,q}(t)$ be the projection
of $O_{p,q} (t)$ mod $L_{-3}$; that is, the expression obtained from
$O_{p,q}(t)$ by setting $L_{-n} = 0$ for $n \geq 3$, with
$L_{-1}$ and $L_{-2}$ taken as commuting.
Then
\eqn{ff2}
\sigma_{p,q}^2 (t) =
{\displaystyle \prod_{{-\frac{p-1}{2} \leq m \leq \frac{p-1}{2}} \atop
                      {-\frac{q-1}{2} \leq n \leq \frac{q-1}{2}} }}
(L_{-1}^2 + 4 (m \theta^{-1} +  n \theta )^2  L_{-2}) \, .
\en

These two results were originally proven using an elegant but rather indirect
argument due to Lutsjuk.\cite{lutsjuk}
Bauer et al. have given\cite{bdfiz2} direct proofs for the cases where
$p=1$ or $q=1$.  They point out, for example, that equation (\ref{bau1})
means that
\eq
f_{p,1} (\lambda, \mu ,t ) =
\det ( - J_- + \sum_{k=0}^{p-1}
        (\mu + \frac{p-1}{2} + J_0 - \lambda k) (-t J_+ )^k )  \, ,
\en
from which a series of $sl(2)$ manipulations gives
\eq
f_{p,1}^2 (\lambda, \mu , t)  =
{\displaystyle \prod_{-\frac{p-1}{2} \leq m \leq \frac{p-1}{2}} }
\left( (\mu + A_{p,1} (m,0))(\mu + A_{p,1} (-m,0)) +
 4 \lambda m^2 t  \right) \, ,
\en
which is equation (\ref{ff1}) with $q=1$.

Now, with the aid of equations (\ref{soln1}-\ref{soln4}), we can complete
these arguments.
We see from equations (\ref{soln1}) and (\ref{soln4}) that
\eqn{rewrite}
O_{p,q}(t)
\prod_{r=0}^{\floor{\frac{q}{2} -1}} \cO_{(q-1-2r)t^{-1} - p ,1} (t) =
\prod_{r=0}^{\ceil{\frac{q}{2} -1}} \cO_{(q-1-2r)t^{-1} + p ,1} (t) \, ,
\en
where for non-commuting operators we take
$\prod_{r=0}^n A_r = A_0 A_1 \ldots A_n$.
Now let us take $t^{-1}$ to be an integer larger than $p$.
Then equation (\ref{soln3}), applied to a vector in $\cFlm$,
implies that
\eqn{rewrite2}
\pi (O_{p,q}(t) )
\prod_{r=0}^{\floor{\frac{q}{2} -1}} \pi ( O_{(q-1-2r)t^{-1} - p ,1} (t) )
w_{n}  =
\prod_{r=0}^{\ceil{\frac{q}{2} -1}} \pi ( O_{(q-1-2r)t^{-1} + p ,1} (t) )
w_{n} \, .
\en
Taking $n = \floor{\frac{q}{2}}p - \floor{\frac{q^2}{4}}t^{-1}$, we obtain
\eq
\begin{array}{rcll}
f^2_{p,q} ( \lambda , \mu , t ) & = &
{\displaystyle \prod_{r=0}^{\ceil{\frac{q}{2} -1}}
f^2_{ p + (q-1-2r)t^{-1},1} (\lambda, \mu - (r+1)(q-r-1)t^{-1} + p (q-r-1), t)}
\times \\
& &
{\displaystyle \prod_{r=0}^{\floor{\frac{q}{2} -1}}
f^{-2}_{ - p + (q-1-2r)t^{-1},1}
(\lambda, \mu - (r+1)(q-r-1)t^{-1} + p (r+1), t) } \, ,
\end{array}
\en
which after some algebra can be reduced to
\eqn{result}
f^2_{p,q} ( \lambda , \mu , t ) =
{\displaystyle \prod_{{-\frac{p-1}{2} \leq m \leq \frac{p-1}{2}} \atop
                      {-\frac{q-1}{2} \leq n \leq \frac{q-1}{2}} }}
\left( (\mu + A_{p,q} (m,n))(\mu + A_{p,q} (-m,-n)) -
 4 \lambda (m \theta^{-1} + n \theta)^2 \right) \, ,
\en
as required.

Since this holds for all integers $t^{-1}$ larger than $p$, and
since both sides are rational functions of $t$,
equation (\ref{result}) must be true for all $t$.
The second result is established similarly.

One further comment: Feigin and Fuchs,\cite{ff3} again following
the method of Lutsjuk,\cite{lutsjuk} have shown
that $O_{p,q} (t)$ is invariant under the transformation
\eq
t \rightarrow -t \, , ~~~~~~~ L_{-k} \rightarrow (-1)^{k-1} L_{-k} \, .
\en
This too can be established by a similar argument.
Let $\widehat{A(t)}$ be the image of an operator $A(t)$ under
the transformation.
It is clear from equation (\ref{bsaops}) that
$\widehat{O_{p,1} (t)} = O_{p,1} (t)$
and $\widehat{O_{1,q} (t)} = O_{1,q} (t)$ for positive integers $p$ and $q$.
{}From equation (\ref{soln3}) it follows that
$\widehat{\cO_{p,1} (t)} = \cO_{p,1} (t)$
and $\widehat{\cO_{1,q} (t)} = \cO_{1,q} (t)$ for any
integers $p$ and $q$.
Now equation (\ref{soln1}) implies that $\widehat{O_{p,q} (t)} =
O_{p,q} (t)$ for any integer $t$; however, since all the coefficients
in these expressions are rational functions of $t$, this establishes the result
in general.

\section{Conclusions}

These simple derivations may perhaps be useful for two reasons.
Firstly, they suggest a general way of deriving analogous results for the
Virasoro algebra's extensions.  Secondly, they illuminate the algebraic
structure generated by the operators $\cO_{a,1}$ and $\cO_{1,b}$, and hint
at the possibility of deriving from first principles the identities,
such as equations (\ref{soln3}) and (\ref{soln4}), which these operators
satisfy.

\acknowledgement

I am very grateful to A. Wassermann and J.-B. Zuber for helpful
discussions.  This work was supported by an SERC Advanced
Fellowship and by the Knox-Shaw Research Fellowship at Sidney Sussex
College, Cambridge.

\end{document}